# Physical mechanisms involved in the formation and operation of memory devices based on a monolayer of gold nanoparticles-polythiophene hybrid materials

*T. Zhang[1], D. Guérin[1], F. Alibart[1], D. Troadec[1], D. Hourlier[1], G. Patriarche[3], A. Yassin[2], M. Oçafrain[2], P. Blanchard[2], J. Roncali[2], D. Vuillaume[1], K. Lmimouni[1], S. Lenfant[1,*]*

[1] Institute of Electronics Microelectronics and Nanotechnology (IEMN), CNRS, University of Lille, Avenue Poincaré, F-59652 Villeneuve d'Ascq, France

[2] MOLTECH-Anjou, CNRS, University of Angers, 2 Bd Lavoisier, Angers, F-49045, France

[3] Centre for Nanoscience and Nanotechnology (C2N), CNRS, University of Paris-Saclay, route de Nozay, Marcoussis, F-91460, France

**ABSTRACT:** Understanding the physical and chemical mechanisms occurring during the forming process and operation of an organic resistive memory device is a major issue for better performances. Various mechanisms were suggested in vertically stacked memory structures, but the analysis remains indirect and needs destructive characterization (e.g. cross-section to access the organic layers sandwiched between electrodes). Here, we report a study on a planar, monolayer thick, hybrid nanoparticle/molecule device (10 nm gold nanoparticles embedded in an electro-generated poly(2-thienyl-3,4-(ethylenedioxy)thiophene) layer), combining, in situ, on the same device, physical (scanning electron microscope, physico-chemical (thermogravimetry and mass spectroscopy, Raman spectroscopy) and electrical (temperature dependent current-voltage) characterizations. We demonstrate that the forming process causes an increase in the gold particle size, almost 4 times larger than the starting nanoparticles, and that the organic layer undergoes a significant chemical rearrangement from a $sp^3$ to $sp^2$ amorphous carbon material. Temperature dependent electrical characterizations of this nonvolatile memory confirm that the charge transport mechanism in the device is consistent with a trap-filled space charge limited current in the off state, the $sp^2$ amorphous carbon material containing many electrically active defects.

## 1. INTRODUCTION

Organic resistive random-access memories (ORRAM) are promising as fundamental elements for the development of electronic organic circuitries[1,2] due to their various advantages, such as simple and low cost fabrication process, low weight, mechanical flexibility and tunable material properties. ORRAM should present two non-volatile resistance states separated by at least 3 orders of magnitude.[1,2] These resistance states can be read by measuring the current at low voltage, and switched by applying a higher voltage sequence.

We have recently demonstrated[3] an hybrid memristive device based on a network of gold nanoparticles (GNP) of 10 nm in diameter connected by chains of a conjugated polymer. This molecules-GNP network was fabricated between two planar platinum electrodes distant of 0.1 to 10 µm. After preparation of GNP functionalized with 10-(2-(3,4-ethylenedioxythiophene)thiophen-3-sulfanyl)decane-1-thiol molecules, the in situ electropolymerization of the thienyl-ethylenedioxythiophene moieties allowed the formation of a polymer-GNP monolayer[4] (see inset in Figure 1b). As in many resistive memories,[1] a forming process is mandatory before operation. The forming process consists of 3 - 4 voltage sweeps between 0 to 20 V at a sweep rate of ~ 4 V/s. Then, the electrical properties of these formed devices (Figure 1) showed systematically (i) a negative differential resistance (NDR) behavior with a peak/valley ratio up to 17 at a voltage of 6-7 V, (ii) a memory effect behavior with an on/off ratio in the $10^3$-$10^4$ range. The retention time for both states was stable during at least 24 h, and a stability of 800 switching cycles between the two states was demonstrated[3]. These 800 cycles between ON and OFF states were observed with a current ratio about $10^3$ without significant degradation. Then, reading pulses up to $10^5$ s were applied to measure the ON and OFF currents. The results showed a good data retention without current drift in this memory.[3] However, no clear physical mechanism was evidenced in this previous work.

Many works on vertically stacked hybrid metal/organic-GNP/metal memories[5;6;7;8;9;10;11;12;13;14;15;16] have propose various mechanisms, such as (i) the formation of metallic filaments[13;14;16;17]; (ii) the electric field-induced charge transfer between organic materials and metallic nanoparticles[6;7]; (iii) the inhibition of charge injection by the space-charge field of trapped charges inducing space charge limited current (SCLC)[8; 12;15] and (iv) the charge trapping/detraping in NPs-organic composites[6] or in NPs associated with a tunneling through organic material[10]. However, these study require destructive analysis, e.g. cross-section to access the hybrid organic-NP layer between the metal electrodes for physico-chemical studies.

In the present work, the advantage of a nanoscale monolayer thick planar structure placed between coplanar electrodes, is that it makes possible to characterize the physico-chemical and electrical structure of the active material on the same device and in operation for a better understanding of the mechanisms responsible for the forming process and the resistive switching memory. Another advantage is the ability to prevent the metal diffusion during the top electrode deposition, which can also flaw the results in vertically stacked devices. Furthermore, the in situ electropolymerization process allows a better control of the organization of the conjugated polymer GNP hybrid material and finally the active layer remains accessible and can be studied by several techniques during voltage cycles in order to better understand the physics of the operation mechanisms. Here, we report on the mechanisms involved in the formation (forming process) and operation of devices based on a network of organic/GNP with a monolayer thickness. For this purpose, different characterizations have been carried out namely (i) the analysis by scanning (SEM) and transmission (STEM) electron microscopies of the morphological changes during the forming process and operation; (ii) the effects of thermal annealing on the morphology of the GNPs network; (iii) the analysis of the organic ligands material at high temperature and (iv) the temperature dependence of the charge transport behaviors in the two-resistance states.

## 2. MATERIALS AND METHODS

**Device fabrication.** *Electrode fabrication.* Devices were processed using a standard electron beam lithography process. We used highly-doped $n^+$-type silicon (resistivity 1-3 mΩ.cm) covered with a thermally grown 200 nm thick silicon dioxide (135 min at 1100 °C in the presence of oxygen 2 L/min followed by a post-oxidation annealing at 900 °C in $N_2$ 2 L/min during 30 min). The planar electrodes were patterned by electronic lithography using 10% MAA 17.5% / PMMA 3% 495K bilayer resists (with thicknesses of 510 nm and 85 nm respectively). Titanium/Platinum, (5/50 nm) were deposited by vacuum evaporation and lift-off. We fabricated electrodes with channel lengths L = 200 nm to 1 μm and channel width W = 100 nm to 1000 μm.

*Functionalized GNPs.* The synthesis of the ligand 10-(2-(3,4-ethylenedioxythiophene)thiophen-3-sulfanyl)decane-1-thiol (**HS-C10-TEDOT**) was described elsewhere[4;18]. The synthesis of 10 nm capped-GNPs (inset in Fig. 1b) involves a ligand exchange by treating 10 nm oleylamine-GNPs[19] by **HS-C10-TEDOT** giving **TEDOT-C10-S-GNPs**. It was previously shown that oleylamine ligands are easily substituted by thiols[20;21;22] (for more details on synthesis see [3]).

*Deposition of the **TEDOT-C10-S-GNPs** monolayer.* Langmuir films of **TEDOT-C10-S-GNPs** were prepared following the method of Santhanam[23] by evaporating a solution of functionalized GNPs in 1,1,2,2-tetrachloroethane on the convex meniscus of a DI (deionized) water surface in a teflon petri dish. The transfer of the floating film on the substrate with lithographed electrodes was realized by dip coating. ***TEDOT-C10-S-GNPs*** form a rather well organized monolayer on the surface (also called NPSAN: Nanoparticle Self-Assembled Network) as shown by SEM image (inset in Fig. 1b). The statistical SEM image analysis gives an average diameter of the GNPs of 9.3 nm, and an average spacing of 2.5 - 3.0 nm between the NPs in the network[3]. Comparison of these distances with the calculated length of 2.1 nm for the free ligand (MOPAC simulation, ChemOffice Software), suggests that the ligand molecules are interdigitated (inset in Fig. 1b). X-ray Photoelectron Spectroscopy (XPS) of the deposited ***TEDOT-C10-S-GNPs*** monolayer on large surface silicon (without electrodes), shows the chemical composition of C10-TEDOT adsorbates before electropolymerization[3]. The good agreement between the theoretical and measured atomic ratios (S2p/C1s) demonstrates the successful grafting of ***HS-C10-TEDOT*** on GNPs[3]. No signal is detected in the N1s region proving the complete substitution of oleylamine ligands by ***HS-C10-TEDOT***.

*Electropolymerization of **TEDOT-C10-S-GNPs** monolayers.* The deposited monolayer of ***TEDOT-C10-S-GNPs*** on coplanar Pt electrodes (inset in Fig. 1b) was electropolymerized in situ, using the Pt lithographed electrodes as working electrodes to form a monolayer film of ***pTEDOT-C10-S-GNPs*** (p stands for polymerized). Electropolymerization of the monolayer was realized in potentiodynamic mode (electrolyte: 0.1M $NBu_4PF_6$ in $CH_2Cl_2$ or $CH_3CN$) by multiple scans at 100 mV/s between -0.4 V and +1 V. This process leads to the development of a broad redox system centered at + 0.7 V vs Ag/AgCl in the cyclic voltammogram (CV). The stabilization of the CV after multiple scans (~ 20) suggests that all redox active TEDOT units have been coupled[4], in agreement with previous results[18].

**Electrochemical experiments** were performed with a Modulab potentiostat from Solartron Analytical. The substrate with lithographed electrodes was hermetically fixed at the bottom of a 0.2 mL Teflon cell containing the electrolyte solution. The counter electrode was a platinum wire of 0.5 mm diameter and Ag/AgCl electrode was used as a reference.

**Electrical measurement setup.** The electrical measurements were performed with an Agilent 4156C parameter analyzer in DC sweeping mode. We used Carl Süss PM5 probe station in order to connect Agilent 4156C and devices. All electrical measurements were performed under inert atmosphere inside a dried-nitrogen filled glove box ($O_2$ < 1 ppm, $H_2O$ < 1 ppm).

**Thermogravimetry coupled with mass spectrometry analysis (TGA/MS).** Thermogravimetry analysis (TGA) (Netzsch STA449F3 Jupiter apparatus) coupled with a quadrupole mass spectrometry (MS) (Aëolos QMS403D, 70 eV, electron impact) through a heated silica capillary system (0.220 mm internal diameter), have been used to monitor the thermal degradation of organic polymers. The TGA provides information relating the changes in sample mass during the annealing process, whereas mass spectrometry offers a method of identifying the evolved volatiles species. Before each experiment, the TGA system was first evacuated and then flushed with ultrahigh purity helium before starting heating. The experiments were carried out under dynamic inert gas atmosphere (helium: 99.999 purity) with a flow rate of 90 $cm^3$/min. The samples were heated using a heating rate of 10°C/min.

**Raman Spectroscopy.** Raman spectroscopy measurements were performed using a Horiba Jobin-Yvon LabRam®HR micro-Raman system combined with a 473 nm laser diode as excitation source focused by a ×100 objective. The scattered light was collected by the same objective in backscattering configuration, dispersed by a holographic grating of 1800 l/mm and detected using a CCD camera.

**Annealing of *pTEDOT-C10-S-GNPs* monolayers.** Freshly prepared *pTEDOT-C10-S-GNPs* monolayers were subjected to a thermal annealing under nitrogen atmosphere (instrument JetFirst 200 from JIPELEC). The annealing temperature (from 423 to 623 K with 50 K steps) was reached in 40 s from ambient temperature, and maintained during 120 s. After annealing, the samples were cooled down to ambient temperature in 180 s.

**Scanning electron microscopy (SEM).** SEM observation was carried out using MEB ultra 55 purchased from Carl Zeiss, or by Focused Ion Beam Strata DB 235 from Fei. The accelerating beam voltage was fixed at 10 kV and 5 kV respectively.

**Scanning transmission electron microscope (STEM) analysis.** Cross-section lamellae for the different samples were prepared with a focused ion beam (FIB) microscope (FEI STRATA DB 235). Milling steps were performed at 30kV with current decreasing from 20 nA to 30 pA while lamella thickness decreased from 6 μm to less than 100 nm, followed by a final step at 5 kV and 30 pA to minimize amorphization. The transfer of the lamella samples to TEM grid was done using a Kleindiek micromanipulator. The STEM observations of these cross-section lamellae were done on a FEI Titan Themis 200 microscope. This microscope is equipped with an aberration corrector on the probe (STEM mode) and the Super-X windowless EDX (4 quadrant SDD EDX detectors with a solid angle > 0.7 srad). The acceleration voltage was 200 kV and the probe current about 50 pA with a half convergence angle of 17.6 mrad. The HAADF-STEM (High Angle Annular Dark Field) images were obtained on the annular dark field detector with a collection angle between 69 and 200 mrad.

## 3. RESULTS AND DISCUSSION

**Evolution of the morphology of *pTEDOT-C10-S-GNPs* monolayer during the forming process.**

After electropolymerization a forming process is mandatory to form the ***formed-pTEDOT-C10-S-GNPs*** monolayer and to "initialize" the device and observe the NDR and memory operation.[3] It consists of 3 - 4 voltage sweeps from 0 to 20 V with a sweep rate of ~ 4 V/s applied to the coplanar electrodes. SEM images of the same device before and after the forming process reveal that the sudden current increase at a voltage of about 6-7 V (see Fig. 4 in Ref. [3]) during the forming process is associate to an irreversible modification of the morphology of the *pTEDOT-C10-S-GNPs* monolayer between the two electrodes (Fig. 2). The SEM image obtained after the forming process (Fig. 2b) exhibits brighter gold clusters between the two electrodes. These clusters can be associated to the reorganization or aggregation, or fusion of the GNPs under the electric field (around 100 MV/m) applied between the electrodes and/or due to the current density induced heating during the forming process. For comparison, the dissolution / nucleation of embedded metallic clusters of platinum inside a $SiO_2$ layer was measured at higher electrical field comprised between 300 – 500 MV/m.[24] Here, the electric field is lower at 100 MV/m and can contribute to a lesser degree to the electro-migration of the GNPs into the polymer. To analyze more precisely these morphological modifications, STEM analysis (see methods) have been performed along the cross-sections of the monolayer (along the dashed line in Fig. 3a and Fig. 3b). The image recorded before the forming process (Fig. 3c) shows GNPs with ~ 10 nm diameter, in agreement with the average value of 9.3 nm determined by the SEM analysis of the monolayer[3]. The image obtained after the forming process (Fig. 3d), clearly shows larger gold clusters with a diameter up to 40 nm. This STEM image of the clusters shows that the clusters have a regular rounded shape rather than a less regular shape expected for aggregation of several GNPs. Moreover, the high-resolution STEM image clearly reveals the monocrystalline structure of this cluster (See Figure SI-1). These observations suggest that the clusters are formed by the melting of several GNPs during the forming process rather than by their aggregation.

**Evolution of the morphology of the *pTEDOT-C10-S-GNPs* monolayer during thermal annealing.**

To understand the physical origin of the cluster formation and to estimate the temperature reached in the device during the forming process, *pTEDOT-C10-S-GNPs* monolayers are annealed at different temperatures from 423 to 623 K with 50 K steps (see methods), and imaged by SEM (Fig. 4). At lower temperatures (423 and 473 K), the GNPs present an average diameter of ~ 10 nm in agreement with the average value of 9.3 nm determine by SEM and STEM analyses of the as-fabricated samples[3]. At 523 K, some clusters with a characteristic size comprised between 20 and 40 nm appear in the monolayer. At higher temperature, 573 K and 623 K, the number of these clusters increases dramatically. Finally, at 623 K, the majority of the GNPs are melted to form these clusters.

The rounded shape and the size of the clusters formed by thermal annealing at 623 K have the same morphology as those observed after the forming process (Figures 2b and 3b). This similarity suggests that during the forming process, the formation of clusters is caused by a local current-induced increase of the temperature in the monolayer (Joule effect).

We observe the beginning of the melting of GNPs at ~ 523 K (Fig. 4) while the melting point (mp) for bulk gold is 1337 K[25]. According to Buffat's thermodynamic model[26;27], which correlates the mp of naked metal nanoparticles to their mean diameter, the mp for 9.3 nm diameter GNP is predicted to be 1273 K. On the other hand, this value is drastically reduced when the GNP surface is coated with molecules[28] or silica[29]. These coating materials increase the surface energy value of the solid phase of the GNPs and reduces mp in agreement with Buffat's model. For example, Miyake et al.[28] have investigated the effect of thermal annealing on 2D superlattices (hexagonal packing) of coated GNPs with dodecane or octadecane-thiol. After thermal annealing at 423 K, the diameter of the dodecane coated GNPs increases from 1.5 ± 0.2 nm to 3.4 ± 0.3 nm. For ~ 10 nm diameter coated GNPs, these authors estimate a mp around 535 K close to the threshold temperature of 523 K observed in our experiments. Thus, we suggest that during the forming process a local temperature of at least of 523 K is reached in the *pTEDOT-C10-S-GNPs* monolayer leading to the formation of clusters by the melting of the 10 nm diameter GNPs.

It was shown that in metal/oxide/metal coplanar structures for memory or memristive applications such as Ag/SiO$_2$/Pt[30], W/SiO$_2$/Ag[24] or Pt/Ag nanoclusters in SiO2/Pt system[31], the formation of metallic clusters may be caused by metal displacement or migration along the direction of the electric field between the metallic electrodes. However, to the best of our knowledge, no previous study mentioning the formation of clusters in organic bistable memories by a thermal effect has been reported so far.

**Thermogravimetry coupled with mass spectrometry analysis (TGA/MS) and Raman spectroscopy.**

Since high temperatures (higher than 523K) can be reach in the samples during the forming process, it was important to assess the thermal stability of organic ligands. Simultaneous thermogravimetry (TG) coupled with mass spectrometry (MS) were used to elucidate the thermal degradation/modification of organics. Two different organic compounds have been tested: i) a conjugated polymer namely *p(TEDOT)* and ii) 2 nm diameter GNPs functionalized with *HS-C10-TEDOT* (insets in Fig. 5). The 2 nm diameter GNPs instead of 10 nm have been chosen, in order to improve the percentage in mass of the organic material, and therefore the sensibility of analytical measurements (TG and MS). Indeed, with a surface density of 4.0 molecules/nm² determined by XPS[3], the percentage in mass of organic material for nanoparticle of 2 nm and 10 nm is estimated to be around 29 % and 8 %, respectively.

Figure 5a and b show the TG/MS traces measured for the 2 nm diameter GNPs functionalized by *HS-C10-TEDOT* ligand (powder was placed directly into the crucible), and for a thin film of *p(TEDOT)* (thickness around 100 µm) on Si/Pt substrate prepared by electropolymerization of *TEDOT* monomers in solution unit[32]. For TG/MS and Raman analyzes on the *p(TEDOT)* film, we use as reference the ***TEDOT*** molecule without the alkylthiol moiety in order to (i) maximize the thickness of the electropolymerized film by removing the electrical resistive part of the molecule and (ii) avoid the oxidation of the thiol group of the molecule during the electropolymerization. We assume that the conclusions on these analyzes on *p(TEDOT)* films are valid for the *p(S-C10-TEDOT)* systems.

The decomposition of ***TEDOT*** moiety in the ***TEDOT-C10-S-GNP*** starts occurring much earlier than *p(TEDOT)* which requires high temperature at about 620 K (Fig. 5b). Both organic-based samples display one major mass loss, between 450 and 720 K for ***TEDOT-C10-S-GNP*** and between 620 K and 720 K for *p(TEDOT)*, involving the release of volatile species corresponding to numerous different ions detected by mass spectrometry. The mass spectrometry spectrum is shown only to emphasize the relationship between the mass losses recorded by TGA and released gaseous species. As the quantity of organics is much lower than the one of gold particles and/or the silicon substrate, the decomposition of organic ligands does not produce a strong enough mass-spectrometry signal for assignment of the series of ions (m/z) to establish a clear-cut fragmentation pathway of evolved volatile molecules. Further work is clearly needed to clarify the decomposition mechanism of these materials.

As with many organic compounds, when ***TEDOT-C10-S-GNPs*** and *p(TEDOT)* are heated, the structural modification leads not only to volatile molecules, but also to a black residue called free carbon. ***TEDOT-C10-S-GNPs*** left a porous brittle black residue unsuitable for further analysis. However, the residue left on the silicon substrate issued from *p(TEDOT)*, easier to handle, has been characterized by Raman spectroscopy. Figure 6 depicts the Raman spectra recorded from the polymer *p(TEDOT)* deposited on Pt/Silicon wafer and the one thermally treated in helium at 940 K. In agreement with the literature[33;34;35], the main Raman bands of ***TEDOT*** moiety are observed in the spectrum and are listed in Table 1. As expected, the heated ***TEDOT-C10-S-GNPs*** -based material exhibited the two characteristic broad Csp2-bands (band D at 1374 cm-1 and band G at 1578 cm-1) of amorphous free carbon.

**Temperature-dependent charge transport mechanisms.**

After the forming process, the device is set in the ON state by applying a double voltage sweep (amplitude from 0 to 20 and 20 to 0 V at 4V/s), and in the OFF state by applying a single voltage sweep (0 to 20 V at 4V/s) followed by an abrupt return to 0V (Fig 1a).[3] The charge transport properties of both states are characterized in the 80 – 300 K temperature range by measuring the ON and OFF current at a fixed reading bias of 1 V (Figure 7). The Arrhenius plots for the ON and OFF currents show that the OFF current is weakly thermally activated with an activation energy of $E_a$ = 51 meV above a threshold temperature of ~ 170 K, and not thermally activated below this temperature. The current in the ON state is not thermally activated in the 80 to 300 K temperature range. This behavior is similar to those previously reported for memory devices based on $TiO_2$ nanoparticles embedded in poly(9-vinylcarbazone) films with no activation energy for the ON state, and an activation energy of ~ 66 meV for the OFF state at temperatures above 150 K.[36]

Several models of charge transport mechanism could be suggested to explain the current-voltage (I-V) characteristics measured on the memory device, such as space charge limited current (SCLC), variable range hopping and thermionic emission limited conduction.[37] I-V characteristics measure at 300 K in the OFF state (Fig. 8) are not consistent with the variable range hopping nor the thermionic emission limited conduction models (for details see Supporting Information). In the case of the SCLC model, I- V curve in log-log scale shows two distinct regions with different I-V relationships (Fig. 8). From low voltages ( < 1 V) in the OFF state the current increases linearly with the bias voltage (Region 1). For voltages between 1 V and 10 V (Region 2), the current exhibits a voltage square dependence. In the ON state, a linear relationship is observed from voltages up to 7 V. The I-V characteristics in the OFF state are well explained by a SLCL mechanism according to the following two equations for low and higher voltage regions respectively (from [37]):

$$I = S\,q\,n\,\mu\frac{V}{s} \quad \text{(low voltage – weak injection)} \quad \text{(Eq. 1)}$$

$$I = S\frac{9\mu\epsilon}{8s^3}\frac{n}{n_t}V^2 \quad \text{(higher voltage – strong injection)} \quad \text{(Eq. 2)}$$

where *I* is current, *q* the electronic charge, *n* density of charge, *µ* mobilité, *V* applied bias, *s* the inter-electrode distance, *e* dielectric constant, $n_t$ density of trapped charge and *S* the electrical contact surface. In the case of the ON state, the current is clearly ohmic. The same behaviors is also observed on shorter inter-electrode gap electrode of 500 nm (see Figure SI-4).

**Physical interpretation of the mechanism in the device**

Cho et al. [36] observed that the ON state of memory devices based on $TiO_2$ nanoparticles embedded in poly(9-vinylcarbazone) films followed an ohmic behavior and a negligible temperature dependence of the current (*i.e.* no thermal activation). They concluded that the current in the ON state is mainly due to charge tunneling between filament conducting paths. Here the same features are observed. Thus, it is likely that the ON state of our device corresponds to a conduction through conducting pathways formed during the "forming process". Based on the above SEM, STEM and Raman measurements, we conclude that the nature of these conducting paths are based on hybrid material composed of amorphous sp2 carbon and gold clusters.

In the off state, we observed a weak thermal activation above 170 K and the I-V characteristics are well explained by a trap-filled SCLC current model. Lin et al.[12] observed also SLCL behavior in the OFF state for devices composed of a layer of polystyrene mixed with GNP of 2-5 nm diameter sandwiched between two aluminum electrodes. As discussed in our previous work[3], a charge carrier trapping/detrapping mechanism from traps can explain these characteristics: (i) the ohmic behavior in ON state corresponds to transport of free carriers in the hybrid material (see above); (ii) the SCLC transport in OFF state corresponds to injection limited transport when charge carriers are trapped in defects (traps) in the materials.

The physical mechanisms proposed here are close to that described by Simmons and Verderber in MIM structures in 1960s, involving charge trapping/detrapping and space charge field inhibition of injection.[38] The nature of the traps involved in the mechanism was generally associated to the NPs present in the hybrid organic/gold nanoparticle memories[3;7;8;10;12]. In the device presented here, after the forming process the resulting material is composed of gold clusters and amorphous sp² carbon. The gold clusters can act as trapping centers but also the amorphous sp² carbon. Sp² carbon was suggested as trapping centers in hybrid poly(N-vinylcarbazole)–graphene nonvolatile memory devices[39]. Regarding the influence of the organic matrix on the electrical properties, the forming process induces an annealing of the organic matrix leading to the formation of amorphous sp² carbon. Moreover, we assume that the observation of switching and clusters formation can occur for other organic polymers incorporated in the device, but with some variation on the voltage values in the memory behavior due to difference of gold diffusivity or sp2 carbon formation for example..

## 4. CONCLUSIONS

Through the course of this study, we followed the physical mechanism involved in the forming process and operations of a hybrid memristive devices based on a network of GNP embedded in an electrogenerated *pTEDOT* matrix. The forming process leads to morphological modifications of the hybrid 2D monolayer film. Larger gold clusters of about 40 nm in diameter appears between the electrodes, similar to those obtained after thermal annealing of the GNP network at ~ 523 K. TG/MS and Raman analysis show evidence of changes in structural organic polymers. The resulting hybrid material, made of amorphous $sp^2$ carbon and gold clusters, is assumed to contain electrical traps, which can explain the voltage and temperature dependence of the charge carrier transport in this device.

## ASSOCIATED CONTENT

**Supporting Information**. This material is available free of charge via the Internet. Cluster analysis by STEM after the forming Process, Model of transport applied on I-V memory characteristic: Variable Range Hopping Model, Thermionic emission limited conduction (TELC) Model and additional I-V curves for shorter devices.

## AUTHOR INFORMATION

**Corresponding Author**

* E-mail: stephane.lenfant@iemn.univ-lille1.fr.

## ACKNOWLEDGMENTS

This work has been financially supported by EU FET project n° 318597 "SYMONE", by the ANR agency, project n° ANR 12 BS03 010 01 "SYNAPTOR" and the French RENATECH network. The authors thank the clean room staff of the IEMN for the assistance and help for the device fabrication.

a)

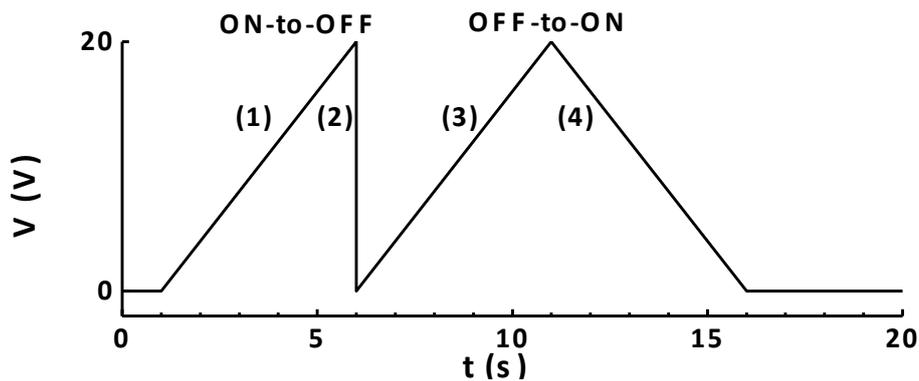

b)

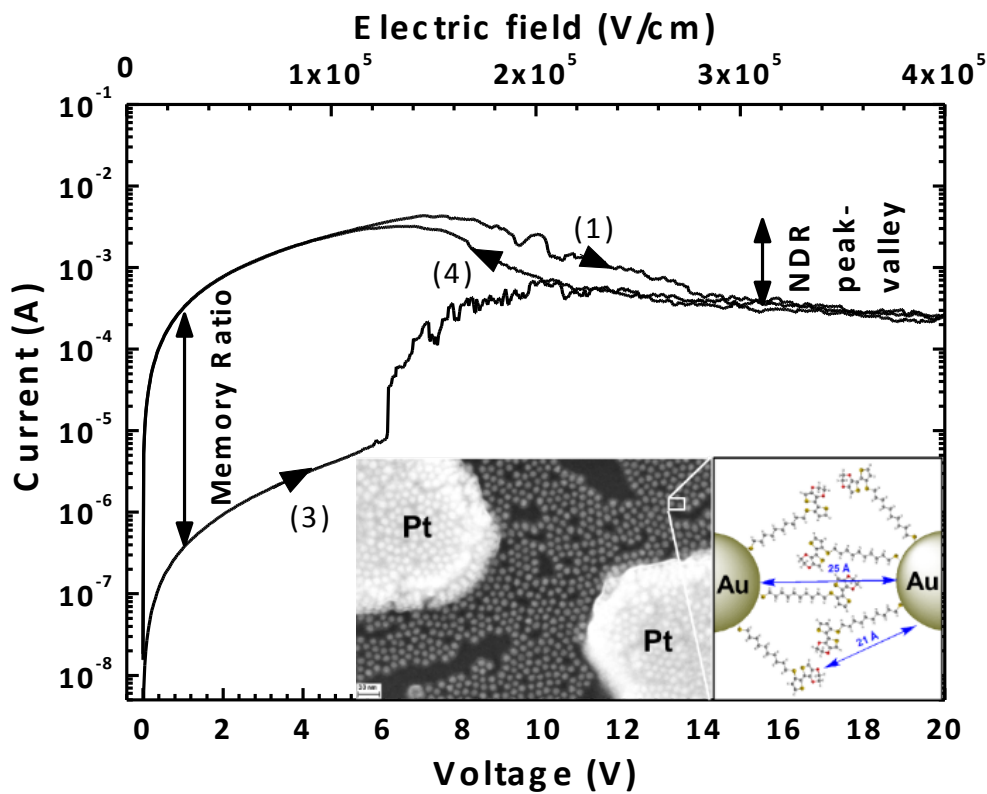

**Figure 1.** **(a)** Typical voltage sequence for the memory behavior and (b) the corresponding I-V curves for the ON-to-OFF and OFF-to-ON switches of a device after the forming process with a length L = 500 nm and width W = 1 mm. The current-voltage I-V traces are numbered according to the voltage sequences shown in (a). Inset, SEM image of a network of functionalized-GNP

monolayer on a nanogap platinum electrodes (electrode spacing L = 200 nm and electrode width W=100 nm); bar scale represents 30 nm; and schematic magnified view of a GNP interspace from [3]. Reprinted from Ref. [3], copyright American Chemical Society 2017.

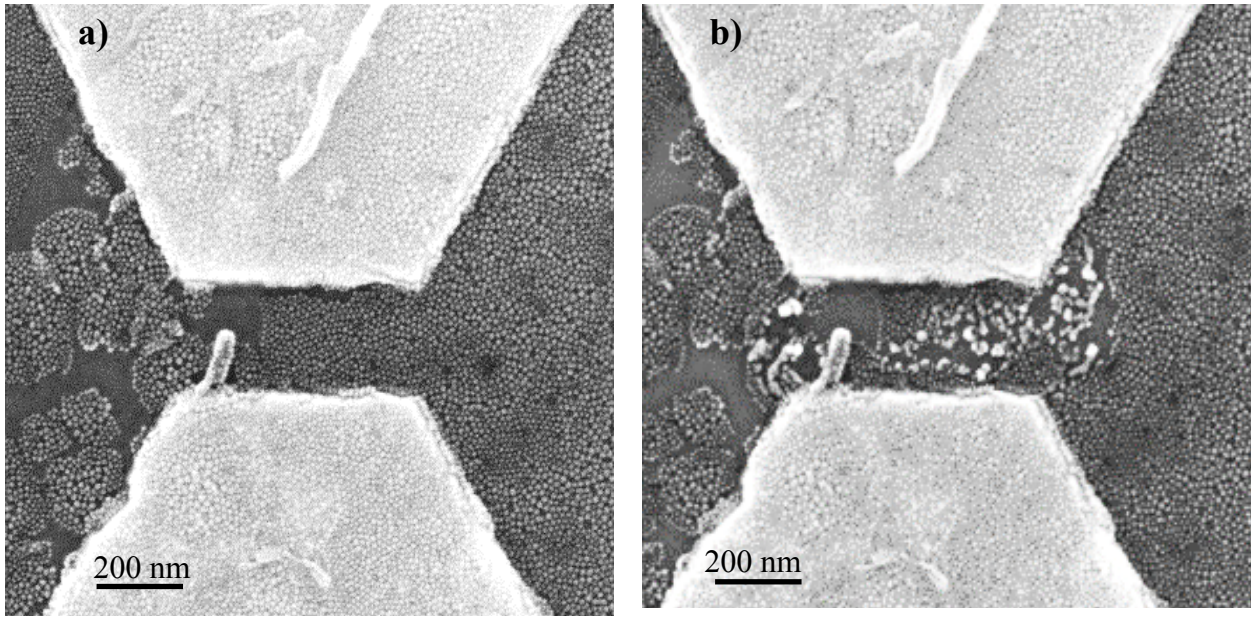

**Figure 2.** SEM images of a *pTEDOT-C10-S-GNPs* monolayer in a 200 nm channel length (GNP size 10 nm) before (a) and after (b) the forming process consisting of 3 - 4 voltage sweeps from 0 to 20 V (corresponding to an electric field maximum of 100 MV/m).

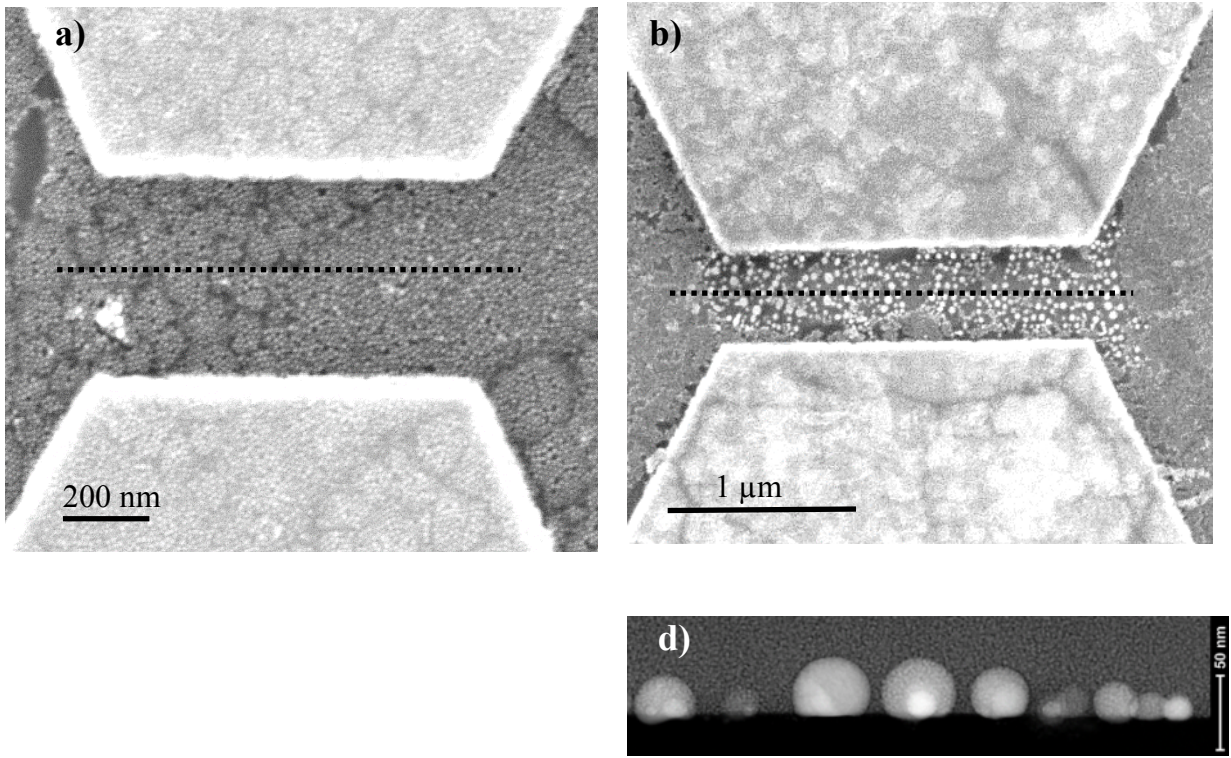

**Figure 3.** Top: (a) SEM image of the *pTEDOT-C10-S-GNPs* monolayer (GNP size 10 nm). (b) SEM image of a *formed-pTEDOT-C10-S-GNPs* monolayer after the forming process. Bottom: corresponding HAADF-STEM images of the cross section between the electrode along the dotted line in SEM images, (c) for device before the forming process (scale bar 10 nm) and (d) after the forming process (scale bar 50 nm). The two devices have the same channel length of 500 nm.

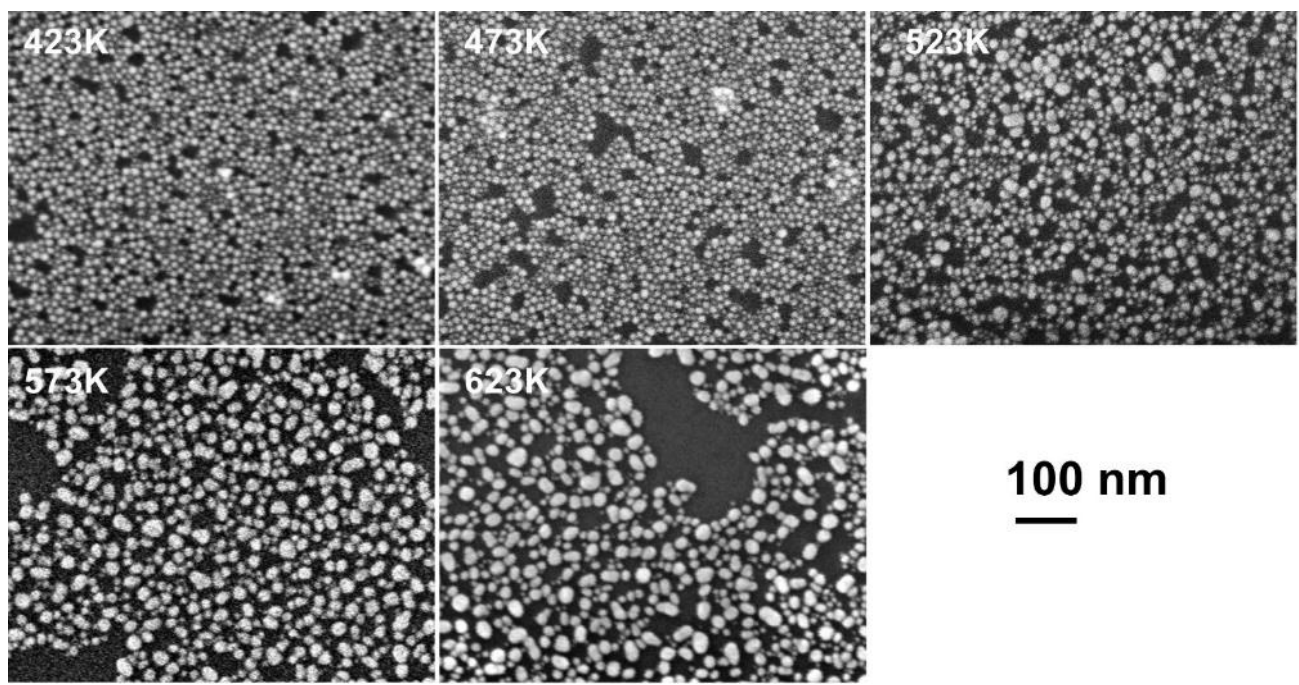

**Figure 4.** SEM images of a *pTEDOT-C10-S-GNPs* monolayer after thermal annealing at different temperatures from 423 to 623 K with 50 K steps. Images were acquired in the gap electrode.

a)

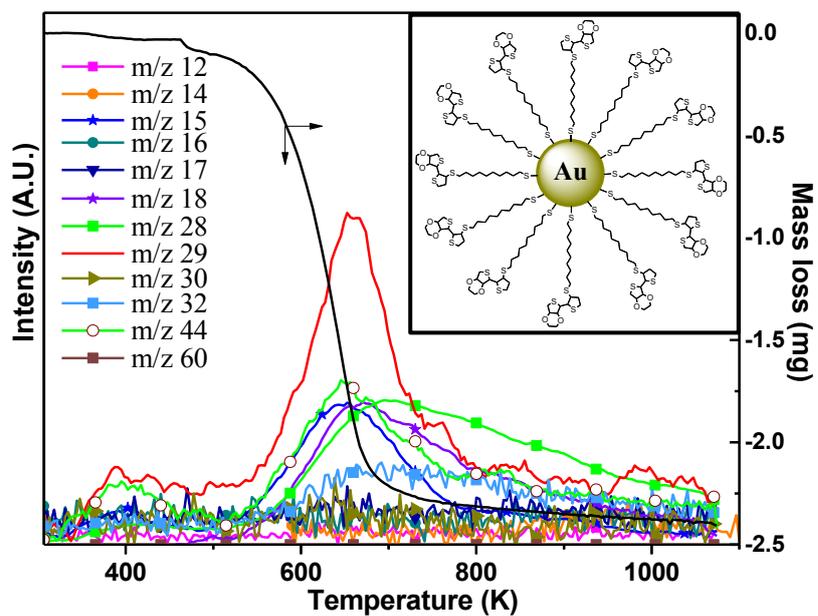

b)

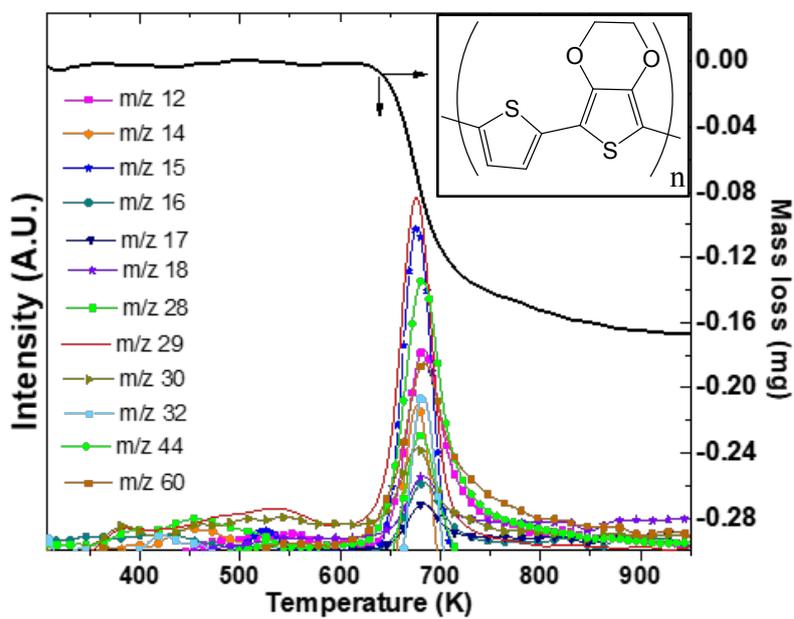

**Figure 5.** Thermogravimetry analysis (TGA) coupled with Mass spectrometry (MS) of (a) a brittle-porous sample of *TEDOT-C10-S-GNPs* with 2 nm diameter GNPs (scheme in the inset) (b) the thin *p(TEDOT)* film on Si/Pt substrate (*p(TEDOT)* structure in the inset).

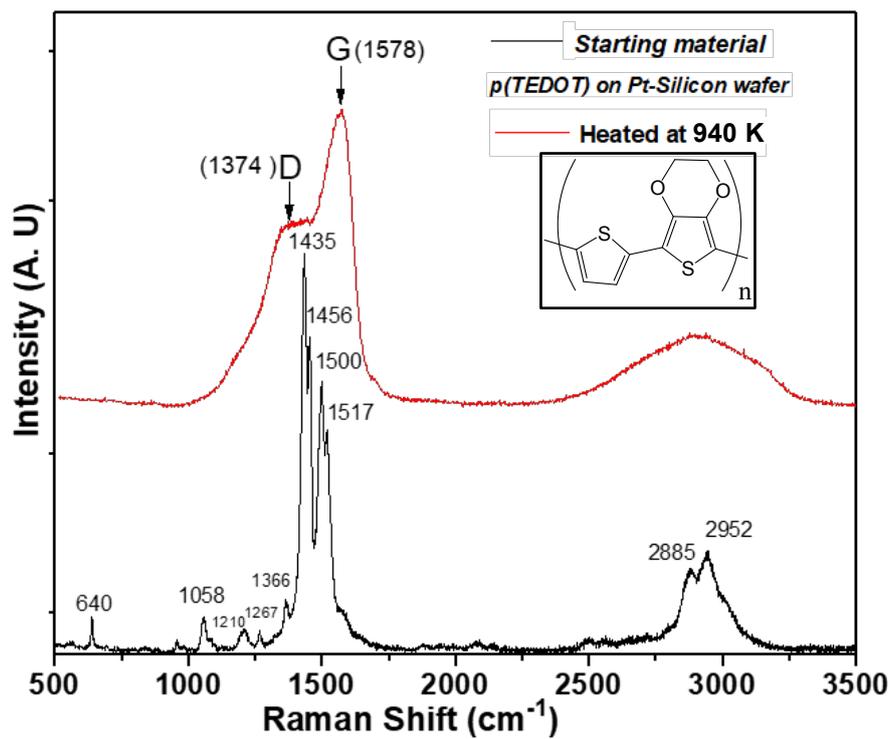

**Figure 6.** Raman spectroscopy acquired from *p(TEDOT)* polymer deposited on Pt-Silicon substrate before and after a thermal annealing in helium at 940 K.

| Wave number (cm-1) | Assignment of Raman bands |
|---|---|
| 640 | Oxyethylene ring deformation |
| 1058 | C-O-C deformation |
| 1210 | stretching $C_\alpha$-$C_{\alpha'}$ (inte-ring) |
| 1267 | stretching $C_\alpha$-$C_{\alpha'}$ (inte-ring), $C_\beta$-H bending |
| 1366 | stretching $C_\beta$-$C_\beta$ intra-ring |
| 1435 | Symmetric stretching $C_\alpha = C_\beta$ (–O) |
| 1456 | Symmetric stretching $C_\alpha=C_\beta$(–H) |
| 1500 | Asymmetric stretching C=C |
| 1517 | Asymmetric stretching C=C |
| 2885 | $CH_2$ Stretching |
| 2952 | $CH_2$ Stretching |

**Table 1.** Characteristic Raman frequencies of *p(TEDOT)* polymer.

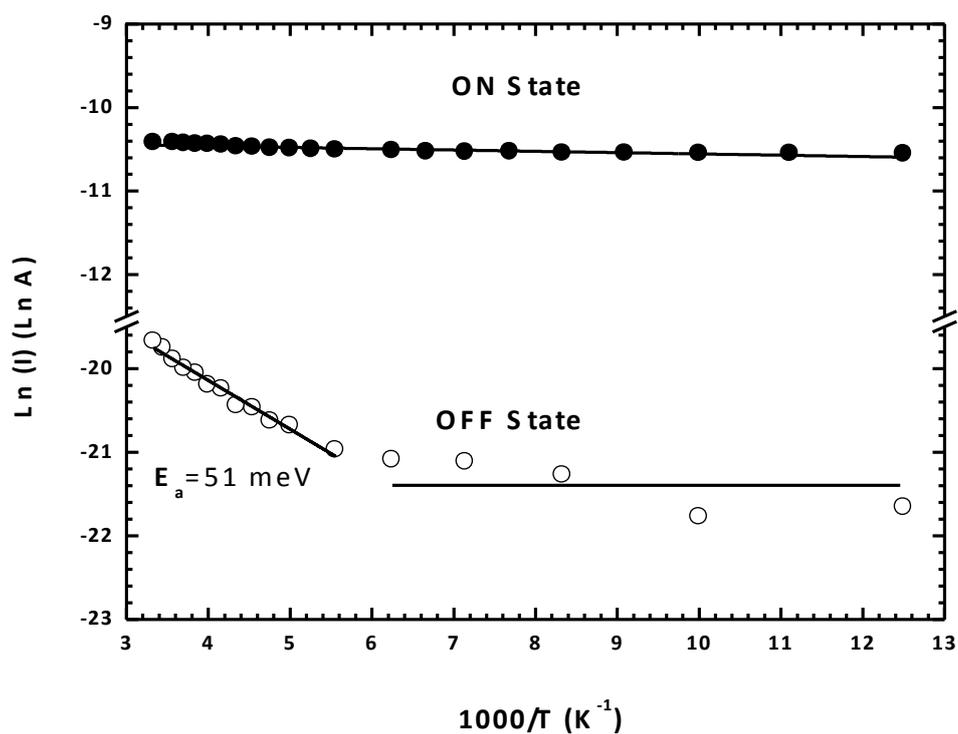

**Figure 7.** Arrhenius plot of the current in the ON and OFF states in the temperature range from 80 to 300 K read at 1 V for a *formed-pTEDOT-C10-S-GNPs* monolayer in a 500 nm x 1000 μm channel device.

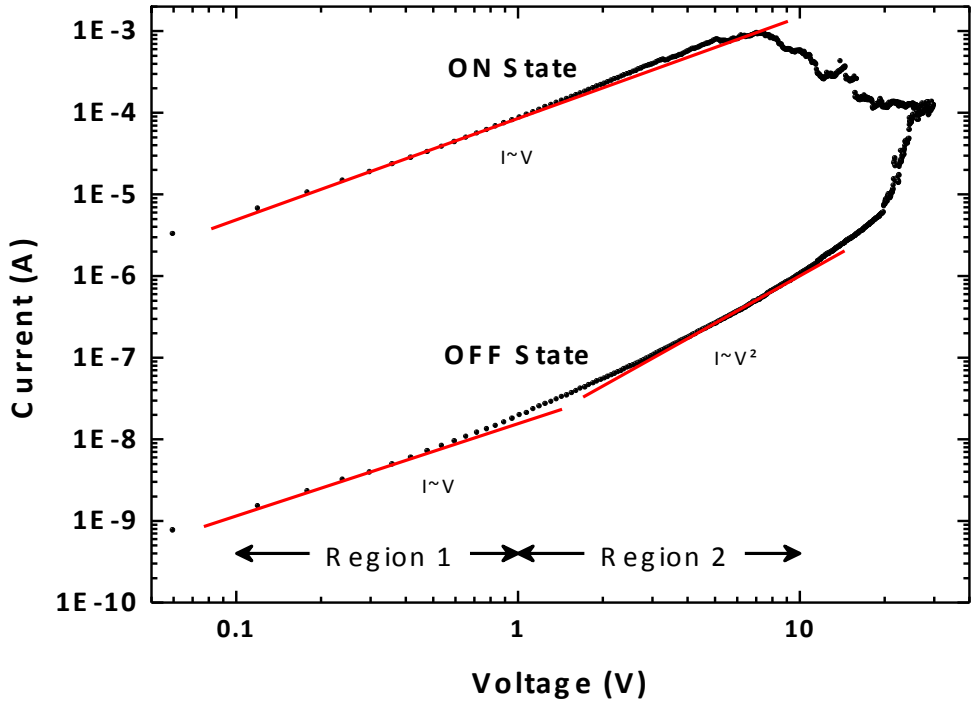

**Figure 8.** Typical current I versus voltage V curves (log-log scale) for the memory behavior measured on a *formed-pTEDOT-C10-S-GNPs* device with a 1 μm gap length. Dots correspond to experimental data and red lines to the fitting adjustments before switching regions; *i.e.* in region 1 (0.1 to 1 V) and Region 2 (1 to 10 V) for the OFF state, and from 0.1 to 7 V for the ON state. From these adjustments at low and high voltage and with Eq. 1 and Eq. 2, we estimate the density of trapped charge $n_t \sim 1.6 \times 10^{14}\,cm^{-3}$.

# Physical mechanisms associated in the formation and operation of memory devices based on a monolayer of gold nanoparticles-polythiophene hybrid materials

T. Zhang[1], D. Guérin[1], F. Alibart[1], D. Troadec[1], D. Hourlier[1], G. Patriarche[3], A. Yassin[2], M. Oçafrain[2], P. Blanchard[2], J. Roncali[2], D. Vuillaume[1], K. Lmimouni[1], S. Lenfant[1]*

[1] Institute of Electronics Microelectronics and Nanotechnology (IEMN), CNRS, University of Lille, Avenue Poincaré, F-59652 Villeneuve d'Ascq, France

[2] MOLTECH-Anjou, CNRS, University of Angers, 2 Bd Lavoisier, Angers, F-49045, France

[3] Centre for Nanoscience and Nanotechnology (C2N), CNRS, University of Paris-Saclay, route de Nozay, Marcoussis, F-91460, France

## SUPPORTING INFORMATION

## 1. Cluster analysis by TEM after the forming Process

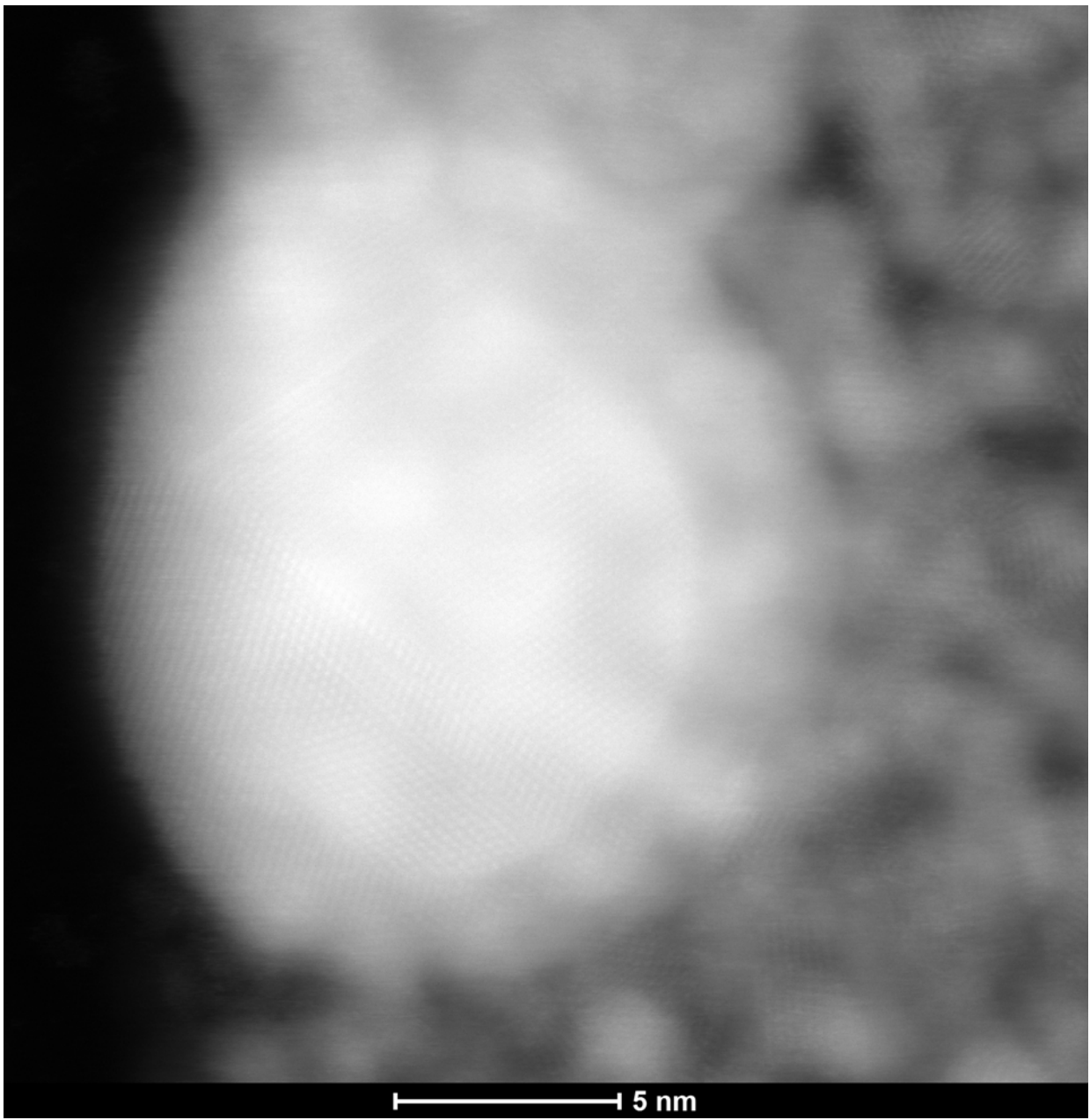

**Figure SI-1.** TEM images of a cluster of the P(TEDOT)-GNPs monolayer after the forming process in the OFF state.

## 2. Model of transport applied on I-V memory characteristics

### 2.1 Variable Range Hopping Model

In the case of a transport mechanism by variable range hopping, the dependence of the conductivity with the temperature is modeled by the following equation (from [1]):

$$\sigma(T) = \sigma_0 e^{\left[-\left(\frac{T_0}{T}\right)^\alpha\right]} \quad \text{(Eq. SI-1)}$$

with $\sigma$ the conductivity, $\sigma_0$ the conductivity prefactor, $T_0$ the characteristic temperature, $T$ the temperature and $\alpha$ a constant equals 1/(D+1) where $D$ is the dimensionality of the electrical conduction path.

---

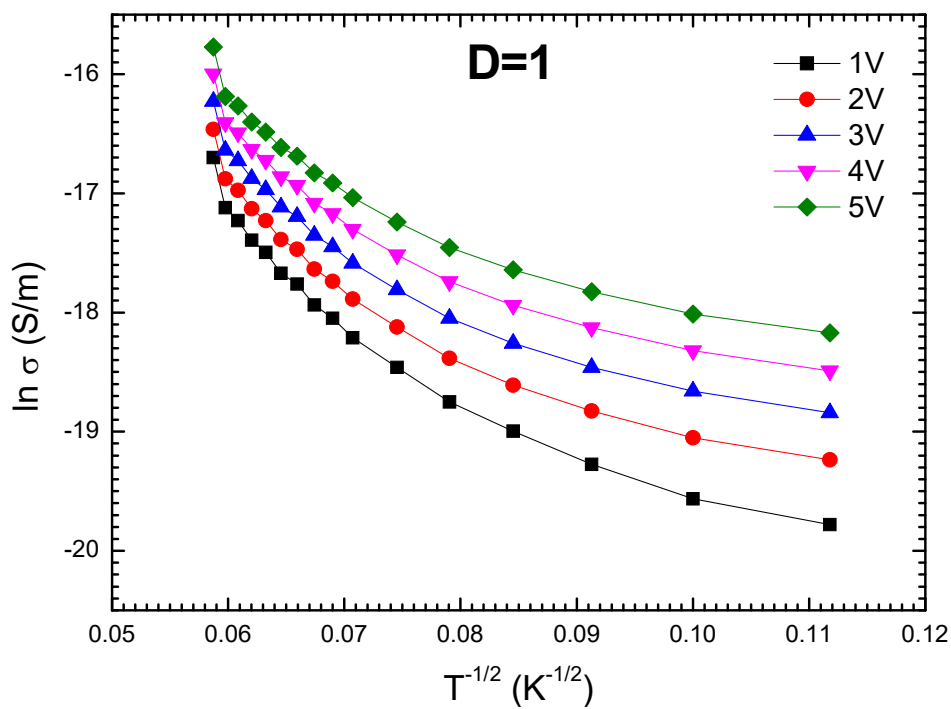

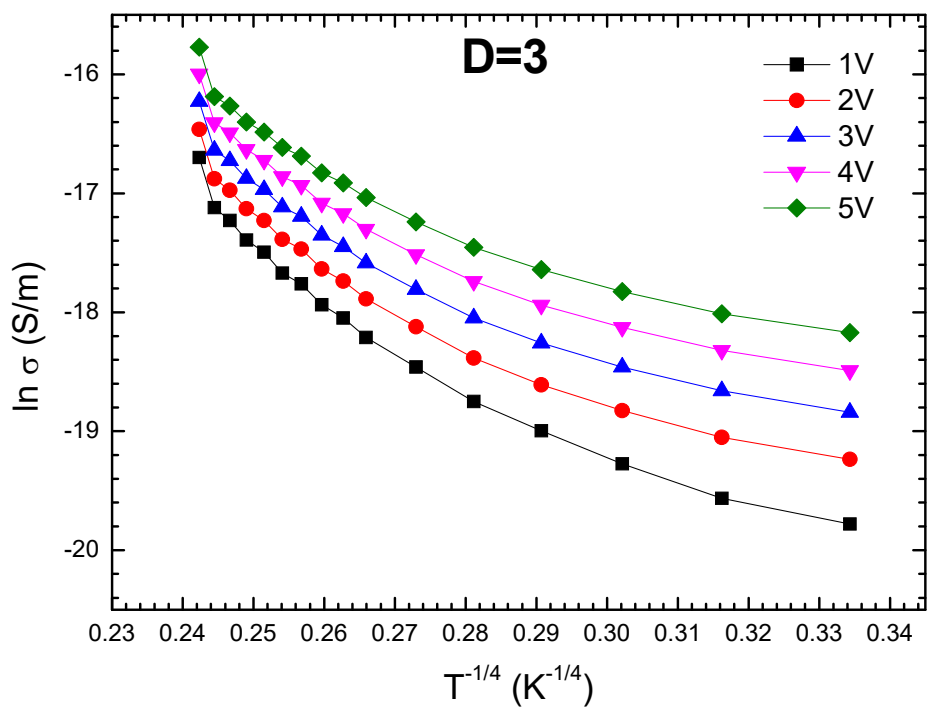

**Figure SI-2.** Temperature dependence of the conductivity measured in a Poly(TEDOT)-GNPs monolayer after the forming process in the OFF state for a 1D and 3D conduction path in top and bottom respectively. No linear behavior is observed. Thus, the transport model by variable range hopping is not consistent with the electrical properties measured on the Poly(TEDOT)-GNPs monolayer after the forming process in the OFF.

## 2.2 Thermionic emission limited conduction (TELC) Model

In the case of a transport mechanism by TELC, the expression of the current is given by the following equation (from [2]):

$$I(V,T) = 120 \frac{m^*}{m_0} T^2 \, S \, e^{-\frac{\phi_0}{kT}} \, e^{\frac{\beta_S E^{\frac{1}{2}}}{kT}} \qquad \text{(Eq. SI-2)}$$

with *J* the current density, *m\** the effective mass of the electron, *m₀* the electron mass, *T* the temperature, *ϕ₀* the barrier height, *k* the Boltzman's constant, *E* the electric field and $\beta_S = \left(\frac{e^3}{4\pi\varepsilon\varepsilon_0}\right)^{1/2}$, with *e* the electronic charge, *ε* permittivity relative, *ε₀* permittivity of vacuum and *S* the cross section.

---

[2] Hesto, P., The nature of electronic conduction in thin insulating layers. In Instabilities in silicon devices, Barbottin, G.; Vapaille, A., Eds. Elsevier: Amsterdam, 1986; Vol. Vol. 1, pp 263-314.

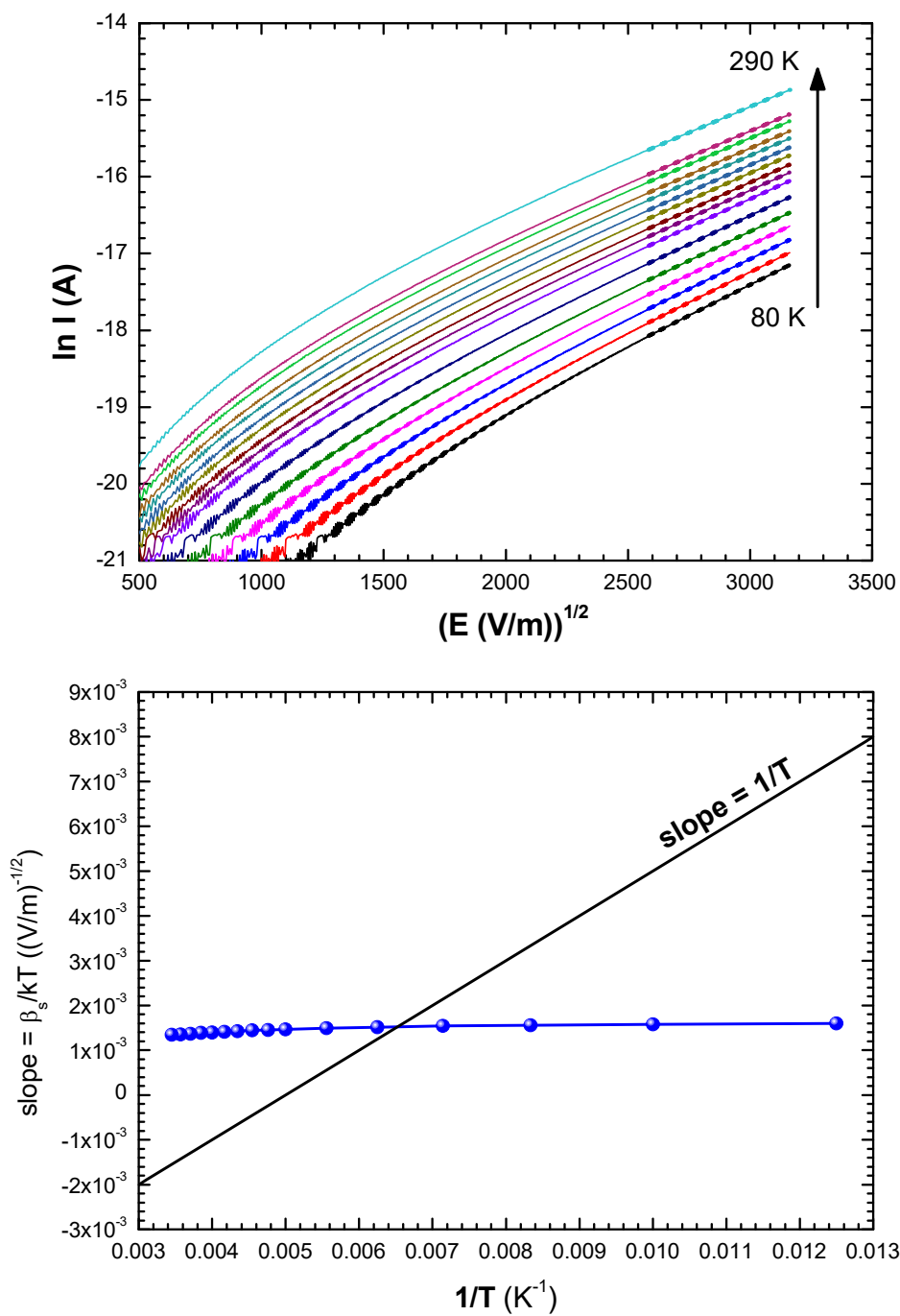

**Figure SI-3.** Top: Electric field dependence of the current, measured in a Poly(TEDOT)-GNPs monolayer after the forming process in the OFF state. From the linear dependence, linear adjustments give the slope of the curves at different temperature. Bottom: the fitted slopes values from the top graphic are plotted in function of 1/T. No linear behavior is observed for this slope with 1/T. Thus, the transport model by TELC is not consistent with the electrical properties measured on the Poly(TEDOT)-GNPs monolayer after the forming process in the OFF.

## 3. I-V memory characteristics on shorter inter-electrode gap

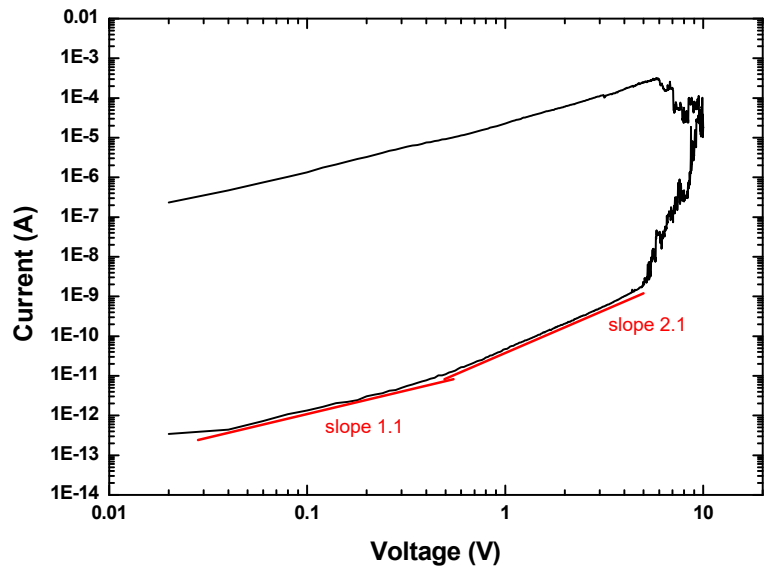

**Figure SI-4.** Typical current I versus voltage V curves (log-log scale) for the memory behavior measured on a formed-pTEDOT-C10-S-GNPs device with a gap length of 500 nm. Dots correspond to experimental data and red lines to the fitting adjustments before switching regions; i.e. in region 1 (0.1 to 1 V) and Region 2 (1 to 10 V) for the OFF state, and from 0.1 to 7 V for the ON state.